\documentclass[a4paper,11pt]{article}
\usepackage{pos}

\title{Quantum Error Correction and $Z(2)$ Lattice Gauge Theories}
\author[a]{Seyong Kim \footnote{In collaboration with M. Rispler (RWTH, Aachen), D. Vodola (INFN, Bologna), and M. M\"{u}ller (RWTH, Aachen)}}

\affiliation[a]{Department of Physics, Sejong University,
05006 Seoul, Republic of Korea}

\emailAdd{skim@sejong.ac.kr}

\abstract{$Z(2)$ lattice gauge theory plays an important role in the study of the threshold probability of Quantum Error Correction (QEC) for a quantum code. For certain QEC codes, such as the well-known Kitaev's toric/surface code, one can find a mapping of the QEC decoding problem onto a statistical mechanics model for a given noise model. The investigation of the threshold probability then corresponds to that of the phase diagram of the mapped statistical mechanics model. This can be studied by Monte
  Carlo simulation of the statistical mechanics model. In~\cite{Rispler}, we investigate the effects of realistic noise models on the toric/surface code in two dimensions together with syndrome measurement noise and introduce the random coupled-plaquette gauge model, 3-dimensional $Z(2) \times Z(2)$ lattice gauge theory. This new Z(2) gauge theory model captures main aspects of toric/surface code under depolarizing and syndrome noise. In these proceedings, we mainly focus on the aspects of Mont Carlo simulation and discuss preliminary results from Monte Carlo simulations of mapped classes of Z(2) lattice theories.}

\FullConference{The 41st International Symposium on Lattice Field Theory (LATTICE2024)\\
 28 July - 3 August 2024\\
Liverpool, UK\\}

\begin{document}
\maketitle

\section{Introduction}

\begin{figure}[ht]
\centering
\includegraphics[width=0.6\textwidth]{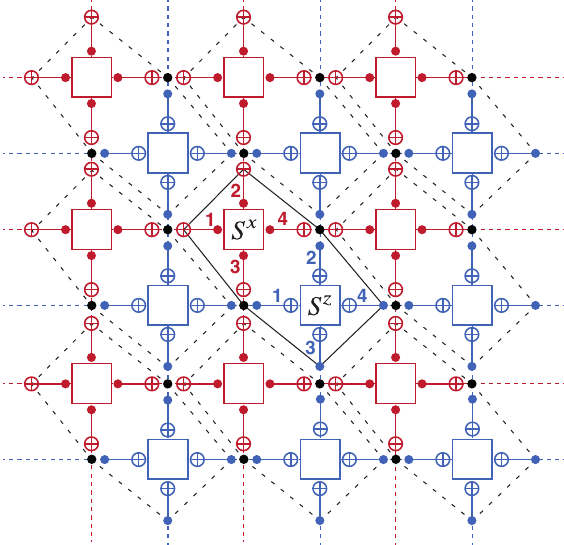}
\caption{The layout for the data and ancilla qubits in
  the toric/surface code. The black dots represent the data
  qubits. The filled colored dot is for the control qubit of the CNOT gate and
  $\bigoplus$ represents the target qubit. Thus the
  red dots represent the ancilla qubits for the $X$-syndrome
  measurements, which acts as the control and NOT operation is
  performed on the data qubit. For the $Z$-syndrome measurements, the
  blue dots is the control for the CNOT gate which coincides with the
  data qubit and NOT operation is performed on the ancilla qubits. A
  syndrome measurement uses four CNOT gates. The number is the CNOT
  operation ordering of these four CNOT gates. Note that the results of syndrome
  measurements are also enumerated according to the two dimensional
  layout of the syndrome measurement circuit. When the syndrome information is not reliable, the syndrome bit can be flipped with a probability, $q$. To deal with this type of noise, syndrome measurements can be repeated adding a discrete time dimension to the model. This entails an error configuration equivalence and can be generalized to so called space-time equivalences. Here, distinct error configurations of qubit and syndrome errors are now possibly equivalent if they produce the same syndrome volume.}
\label{fig:qubit_schematic}
\end{figure}

For certain classes of computational problems (e.g.~\cite{Shor,
  Funcke:2023jbq}), Quantum Computers (QC) are expected to outperform
digital supercomputers through superposition and entanglement of
quantum states due to the parallelism inherent in these quantum
states. Difficult problems in lattice QCD such as real time phenomena
and QCD in finite baryon density may benefit from the
development of QC~\cite{Funcke:2023jbq}. Constructing QC's which are competitive enough
against digital supercomputers is challenging and may take decades
to develop, and currently it is actively being pursued by many
researchers (e.g., ~\cite{google_quantum}).

To build a quantum computer, the ability to keep the qubit states stable and to change the qubit states reliably are basic
requirements. Scaling up this ability to a system with a large number of qubits is also necessary~\cite{DiVincenzo:2000}. Similar to bits in digital computer, qubits are susceptible to environmental effects, which leads to qubit errors. The
control of qubits is a quantum mechanical evolution of the qubit
system and is also susceptible to environmental effects. In the
foreseeable future, limiting these sources of errors to a negligible
amount will be difficult and a fault-tolerant quantum computer
architecture shall be essential in constructing quantum
computers\footnote{Note that in a modern digital computer, error
  detection and correction is already an essential part of
  the architecture (e.g., such as ECC RAM~\cite{ECC_RAM}) and that most
  digital computers do work fault-tolerantly in noisy environment when
  the noise level is low.}. Implementing Quantum Error Correction
(QEC) and error detection in quantum memory and quantum
gate operation is a first step to build and operate a quantum computer
in a fault-tolerant way~\cite{Preskill:1997ds,Campbell:2017}.

Like the Hamming code in digital computer~\cite{Hamming}, typical
ideas for QEC such as the toric/surface code and topological color codes rely on forming few logical qubits out of many physical qubits. A practical
question would be ``how many physical qubits do we need for a logical
qubit?'' and notably the answer depends on the implemenation of the QEC protocol and the underlying noise model. %the answer depends on the quantum error of underlying
physical qubits and physical gates which form a QEC protocol and its implementation. 
In this
work, we investigate this problem through a mapping of this question
to a statistical mechanics model where successful decoding for QEC can be identified with the existence of an 'ordered phase' and Monte Carlo (MC)
simulation of the mapped model (see, e.g., ~\cite{Dennis:2001nw} and
~\cite{Vodola:2021hsy} for some examples). A maximum allowed quantum
error rate in a successful QEC protocol, called threshold
probability, is related to the disorder probability of couplings (or
wrong-sign couplings) and the temperature in a spin model. We specifically discuss a class
of the statistical mechanics models corresponding to the decoding of
quantum error patterns which may occur in the quantum circuits
implementation of the toric/surface code in
Fig.~\ref{fig:qubit_schematic} (the toric/surface code still is one
of the strong candidates for realizing fault-tolerant quantum
computing among many encoding algorithms despite decades of
researches on the topic). Quantum errors assumed in this work are Pauli
errors ($X$-error or bit-flip error, $Z$-error or phase-flip error,
$Y$-error or bit-flip together with phase flip error) and syndrome
measurement errors (i.e., errors in the quantum circuit for the
ancilla qubit measurements).

In contrast to the threshold probability obtained with leading practical decoding algorithms such as Minimum-Weight Perfect Matching (MWPM), our MC
simulation results  give higher threshold probability
(see~\cite{Rispler} for detail), which implies that there is a
room for improvement in quantum error decoding algorithm
development. % MC result also suggests the upper bound on the quantum error rates that hardware develop needs to achieve in order to build a fault-tolerant quantum computer. 

\section{Method}

The mapping to a statistical mechanics model proceeds with analyzing quantum
error patterns in Fig.~\ref{fig:qubit_schematic}. In principle, Pauli
errors are assumed to happen in any of data qubits and
ancilla qubits and the errors in one-qubit gates and two-qubit gates
(CNOT gates) can also occur. Assigning
the same error probability $p$ to qubit errors and considering how the
error propagates through CNOT gates and simplifying the possibilities~\cite{Rispler},
we arrive at
\begin{align}
{\rm pr}(X_h) = \frac{8p}{3}  + \frac{48p}{15}, \;\;\;\;\; 
{\rm pr}(X_{\rm v}) = \frac{8p}{3}  + \frac{32p}{15}, \;\;\;\;\;
{\rm pr}(q) = \frac{8p}{3}  + \frac{48p}{15}
\label{prob_RPM}
\end{align}
for the case of bit-flip error, ${\rm pr} (X)$ (the subscript $h$ and
${\rm v}$ denotes horizontal and vertical direction
respectively), plus syndrome error ${\rm pr} (q)$, or for the case of
phase-flip error (${\rm pr} (X)$ is renamed as ${\rm pr} (Z)$) plus
syndrome error, ${\rm pr} (q)$, and
\begin{align}
{\rm pr}(X_h) = \frac{4p}{3}  + \frac{32p}{15}, \;\;\;\;\;
{\rm pr}(X_{\rm v}) = \frac{4p}{3}  + \frac{16p}{15}, \;\;\;\;\;
{\rm pr}(Y_h) = \frac{4p}{3}  + \frac{16p}{15} = {\rm pr}(Y_{\rm v})\\
{\rm pr}(Z_h) = \frac{4p}{3}  + \frac{16p}{15}, \;\;\;\;\;
{\rm pr}(Z_{\rm v}) = \frac{4p}{3}  + \frac{32p}{15}, \;\;\;\;\;
{\rm pr}(q) = \frac{8p}{3}  + \frac{48p}{15}
\label{prob_aniso}
%\MR{{\rm pr}(\tilde q) = \frac{10p}{3}  + \frac{48p}{15}.}
\end{align}
for the case of $X, Y$ and $Z$ error plus syndrome error. The so-called
Nishimori line (the red dotted line in Fig.~\ref{fig:Phase}) is the
line given by 
\begin{equation}
  \exp{(- 4 |J (W)|)} = \frac{{\rm pr}(X) {\rm pr}(Y) {\rm pr}(Z)}{{\rm
      pr}(W)^2 {\rm pr}(I)}, \;\;\; \exp{( - 2 {|{J_t}^{\sigma, \tau}|})} =
  \frac{q}{1-q}, \label{nishimori}
\end{equation}
through the condition that the disorder distribution is equal to the
thermal distribution~\cite{Nishimori2002}.

\begin{figure}[ht]
\centering
\begin{minipage}[b]{0.45\linewidth}
\includegraphics[width=\textwidth]{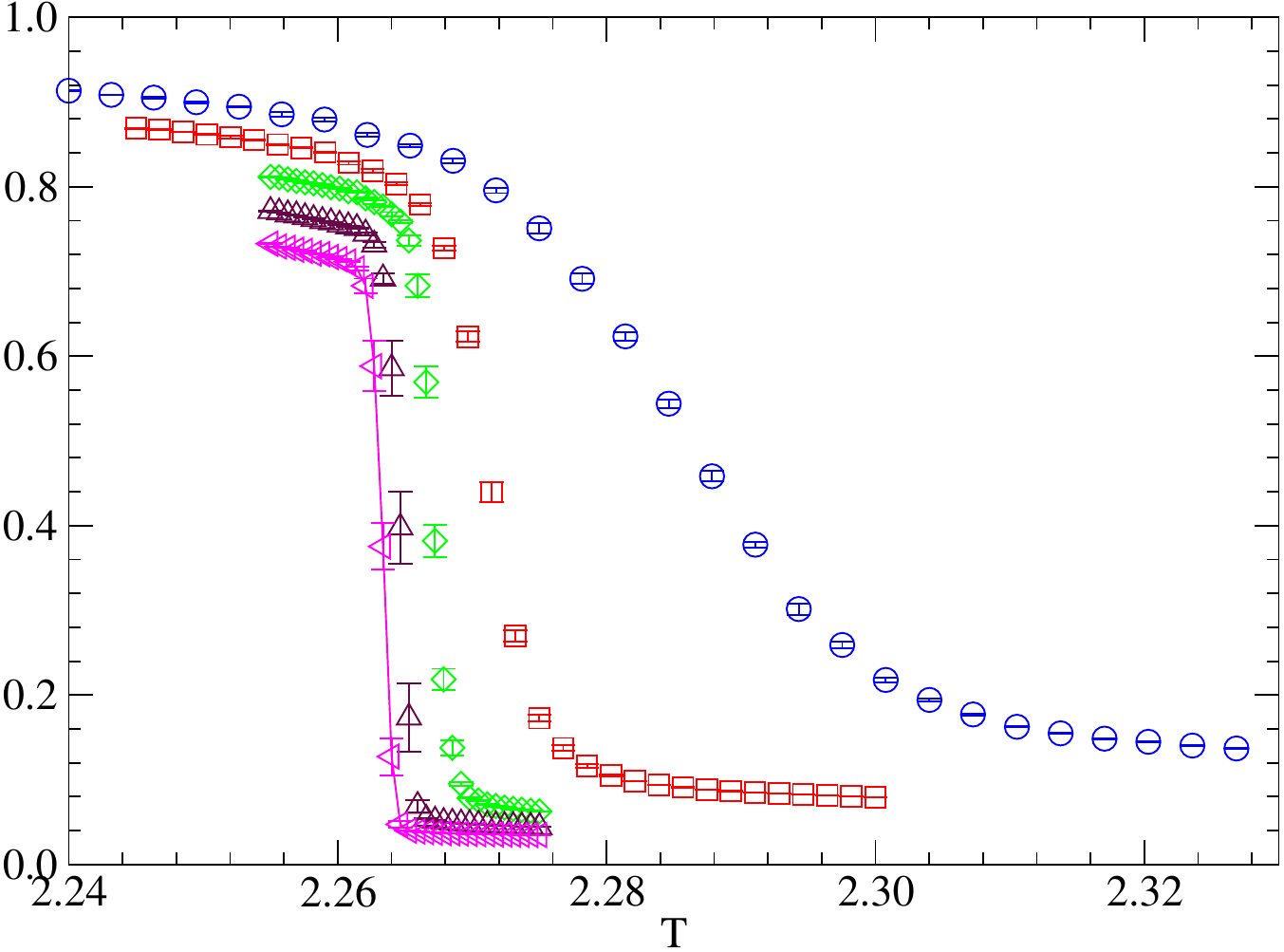}
\end{minipage}
\quad
\begin{minipage}[b]{0.45\linewidth}
\includegraphics[width=\textwidth]{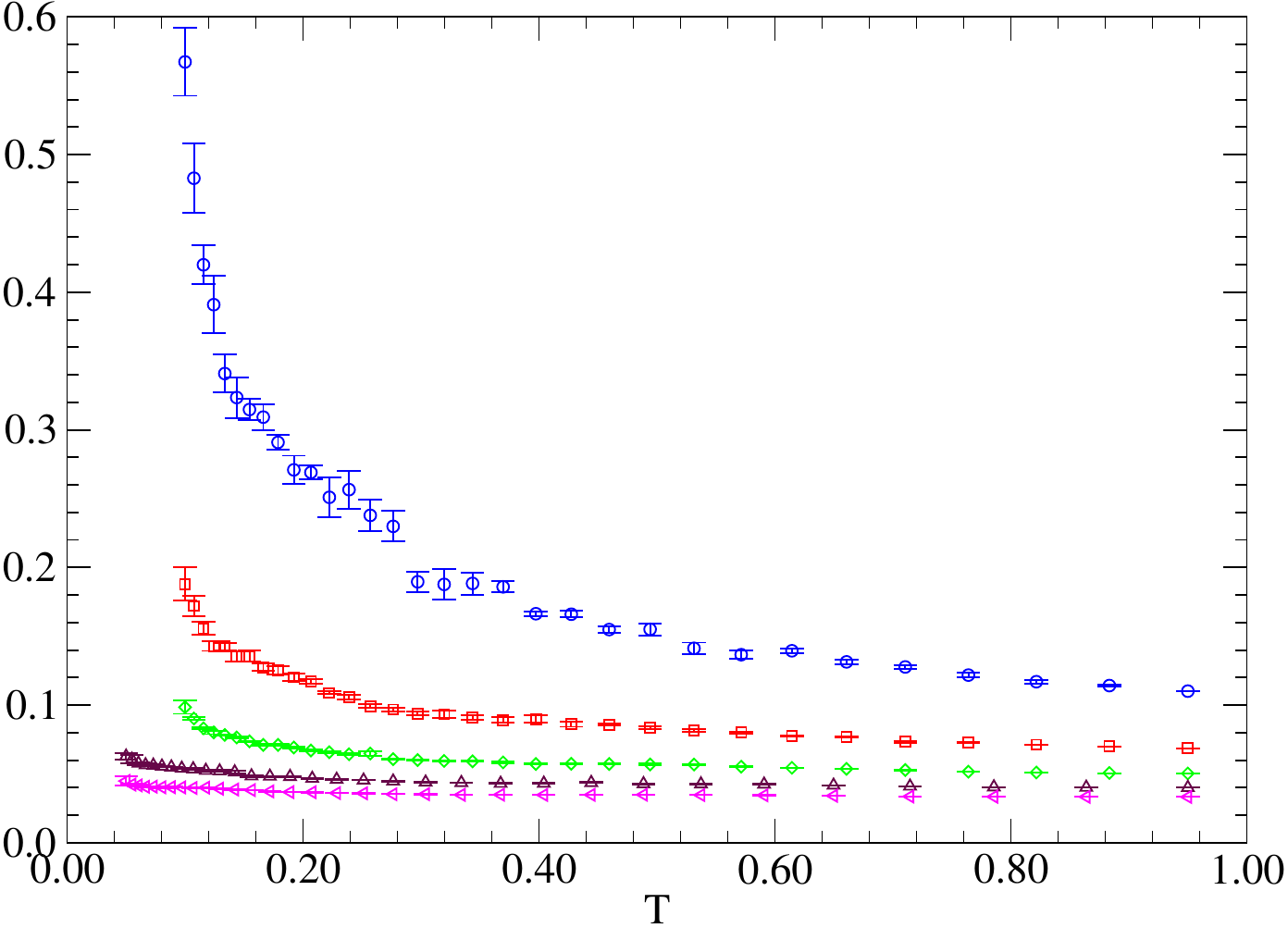}
\end{minipage}
\caption{The Polyakov line at $p = 2.88 \times 10^{-5}$ (left) and
  $p = 2.31 \times 10^{-2}$ (right) for the circuit level
  noise. MC Simulation of $Z(2) \times Z(2)$ gauge theory on $8^3$
  (blue circle), $12^3$ (red square), $16^3$ (green diamond), $20^3$
  (maroon up-triangle), and $24^3$ (magenta left-triangle). From
  Eq.~\ref{prob_aniso}, ${\rm pr} (X_h) = \frac{52p}{15} = 0.0001$
  (left) and $0.08$ (right).} 
\label{fig:Poly}
\end{figure}

\begin{figure}[ht]
\centering
\begin{minipage}[b]{0.45\linewidth}
\includegraphics[width=\textwidth]{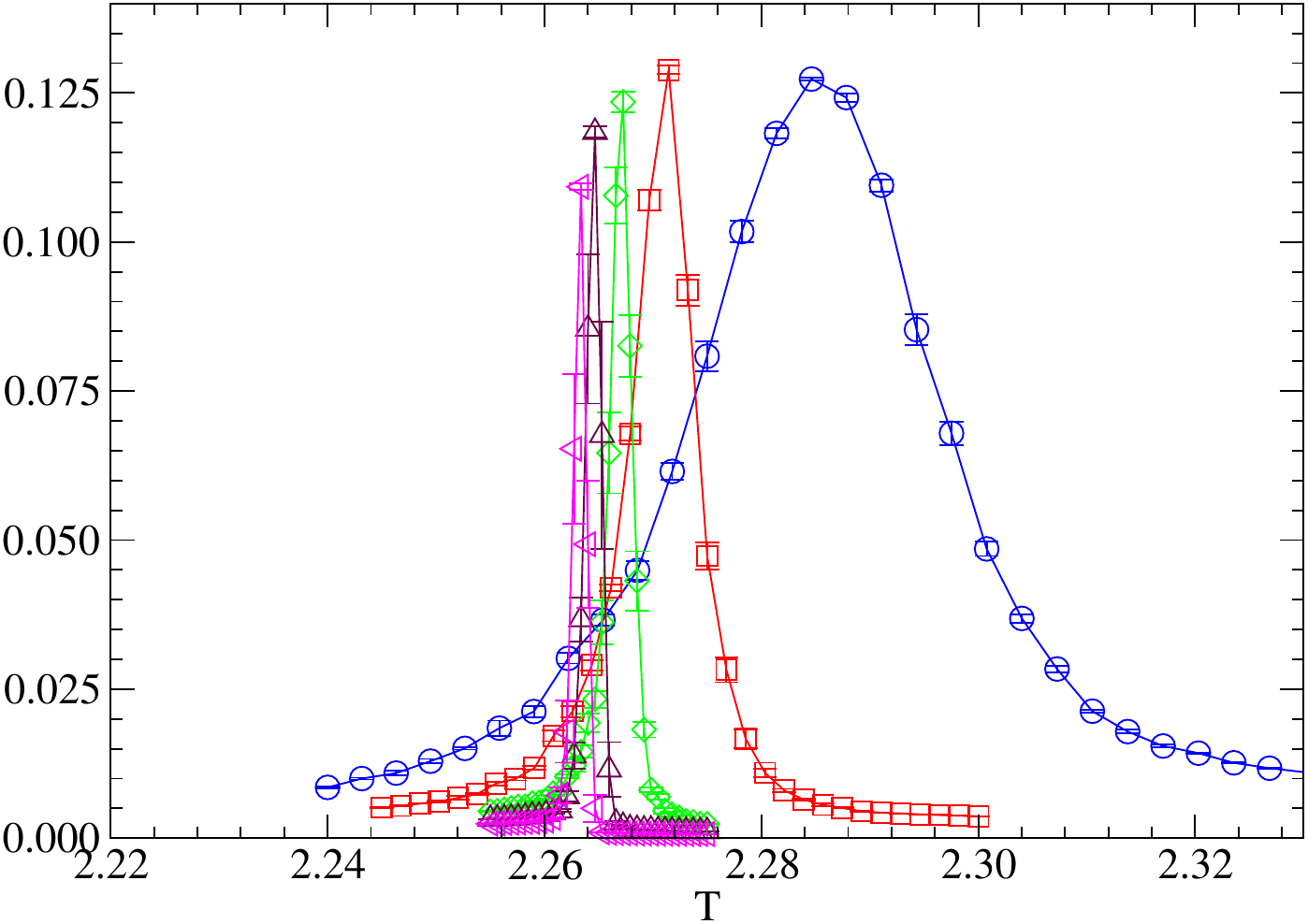}
\end{minipage}
\quad
\begin{minipage}[b]{0.45\linewidth}
\includegraphics[width=\textwidth]{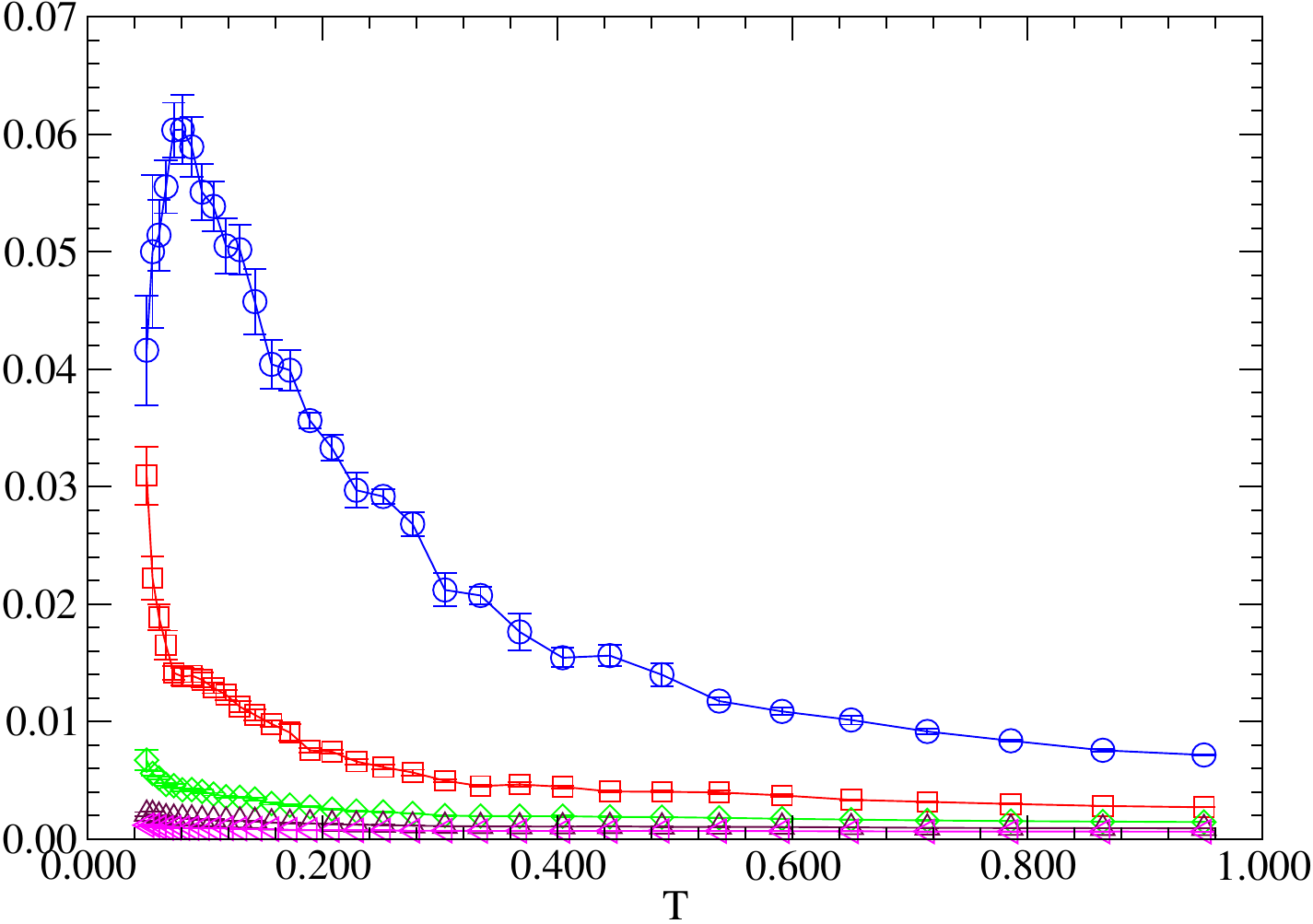}
\end{minipage}
\caption{The Polyakov line susceptibility at $p = 2.88 \times 10^{-5}$ (left) and
  $p = 2.31 \times 10^{-2}$ (right). Symbols and colors are the same
  as in Fig.~\ref{fig:Poly}.}  
\label{fig:Polysusc}
\end{figure}

\begin{figure}[ht]
\centering
\begin{minipage}[b]{0.45\linewidth}
\includegraphics[width=\textwidth]{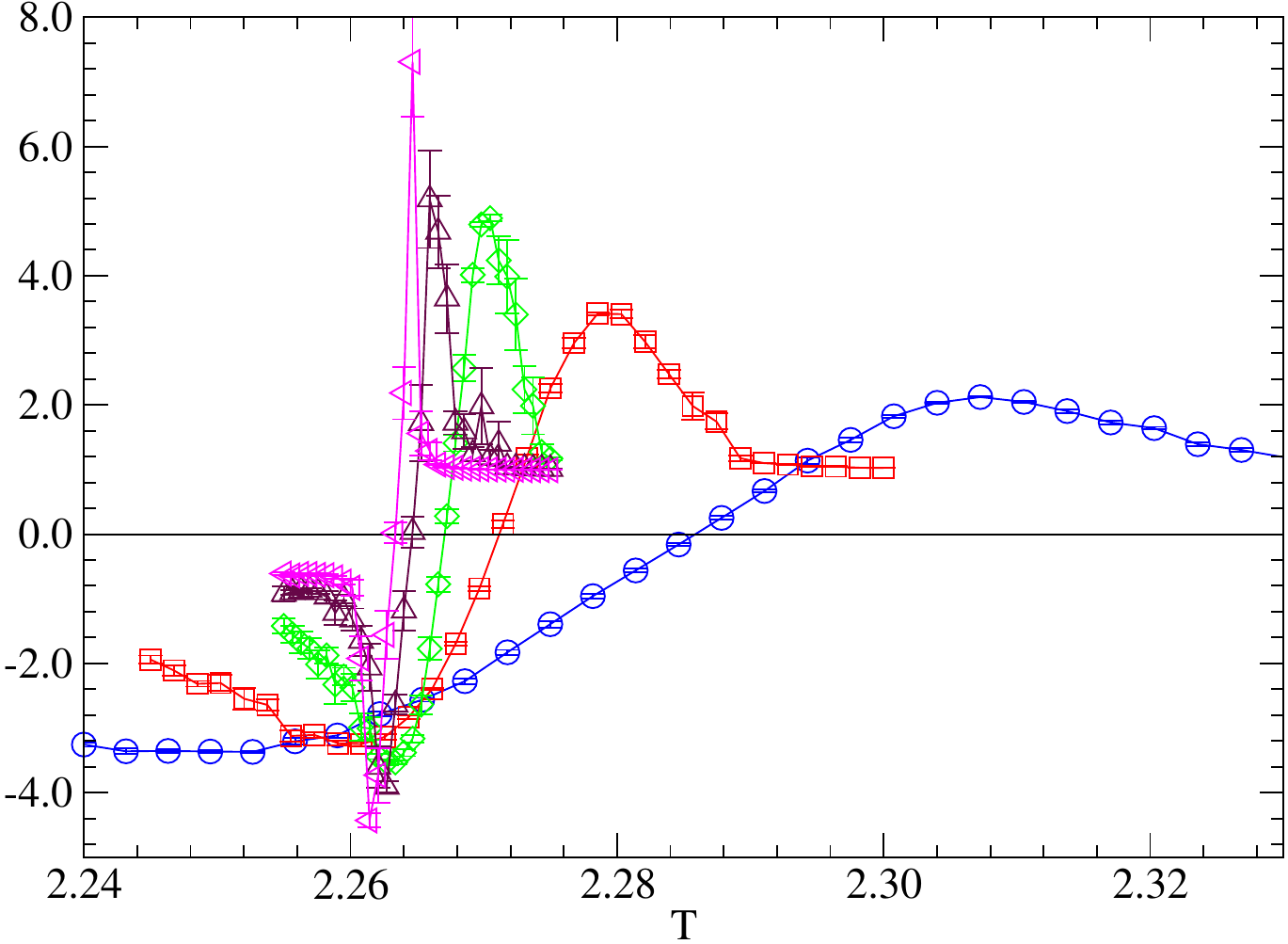}
\end{minipage}
\quad
\begin{minipage}[b]{0.45\linewidth}
\includegraphics[width=\textwidth]{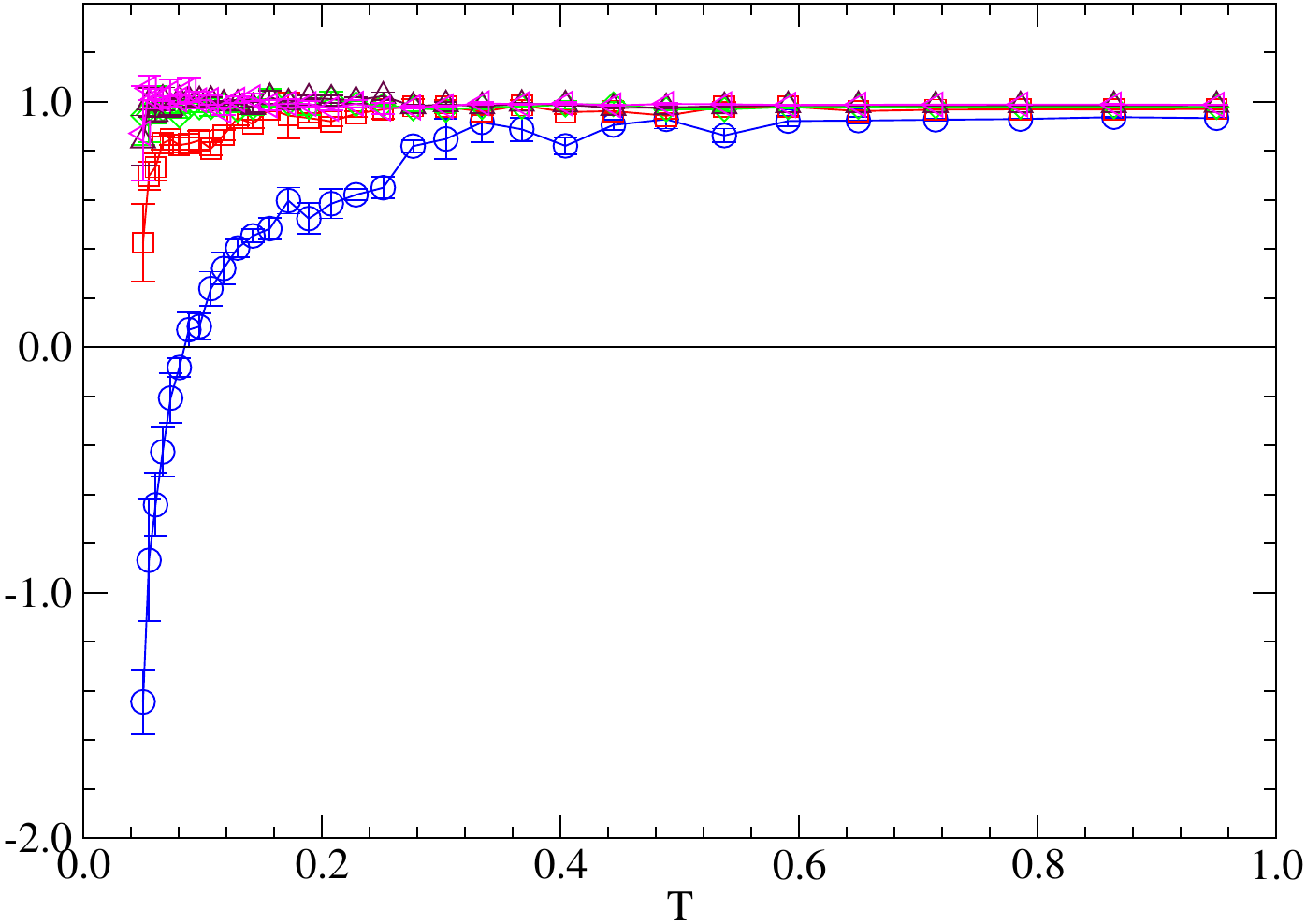}
\end{minipage}
\caption{The third order cumulant of the Polyakov line at $p = 2.88 \times 10^{-5}$ (left) and
  $p = 2.31 \times 10^{-2}$ (right). Symbols and colors are the same
  as in Fig.~\ref{fig:Poly}.}
\label{fig:B3}
\end{figure}

Then, the mapped statistical mechanics models can be described by the Hamiltonian
\begin{equation}
H = \sum_{\mathbf{n}} \left[ H_X (\mathbf{n}) + H_Y (\mathbf{n}) + H_Z (\mathbf{n}) \right],
\label{eq:Hamiltonian}
\end{equation}
where
\begin{align}
H_X (\mathbf{n}) = 
- {J_x}(\mathbf{n};X) \sigma_y(\mathbf{n}) \sigma_t(\mathbf{n} + \hat{y})
\sigma_y(\mathbf{n}+\hat{t}) \sigma_t(\mathbf{n}) 
- {J_y}(\mathbf{n};Y) \sigma_t(\mathbf{n}) \sigma_x(\mathbf{n}+\hat{t}) 
\sigma_t(\mathbf{n}+\hat{x}) \sigma_x(\mathbf{n}) \nonumber \\
- {J_t}^\sigma (\mathbf{n};q) \sigma_x(\mathbf{n}) \sigma_y(\mathbf{n}+\hat{x})
\sigma_x(\mathbf{n}+\hat{y}) \sigma_y(\mathbf{n}) \nonumber \\
H_Z (\mathbf{n}) = 
- {J_x}(\mathbf{n};Z) \tau_y(\mathbf{n}) \tau_t(\mathbf{n}+\hat{y})
\tau_y(\mathbf{n}+\hat{t}) \tau_t(\mathbf{n}) 
- {J_y}(\mathbf{n};Z) \tau_t(\mathbf{n}) \tau_x(\mathbf{n}+\hat{t})
\tau_t(\mathbf{n}+\hat{x}) \tau_x(\mathbf{n}) \nonumber \\
- {J_t}^\tau (\mathbf{n};q) \tau_x(\mathbf{n}) \tau_y(\mathbf{n}+\hat{x})
\tau_x(\mathbf{n}+\hat{y}) \tau_y(\mathbf{n}) \nonumber \\
\end{align}
for the $X$-error (bit flip) and the $Z$-error (phase flip) and
\begin{align}
H_Y ({\mathbf{n}}) = 
-{J_x}(\mathbf{n};Y) \sigma_y (\mathbf{n}) \sigma_t(\mathbf{n}+\hat{y})
\sigma_y(\mathbf{n}+\hat{t}) \sigma_t(\mathbf{n}) 
\tau_t(\mathbf{n}+\hat{x}) \tau_x(\mathbf{n}+\hat{x}+\hat{t})
\tau_t(\mathbf{n}+\hat{x}+\hat{x}) \tau_x(\mathbf{n}+\hat{x}) \nonumber \\ 
- {J_y}(\mathbf{n};Y) \sigma_t(\mathbf{n}) \sigma_x(\mathbf{n}+\hat{t})
\sigma_t(\mathbf{n}+\hat{x}) \sigma_x(\mathbf{n}) 
\tau_y(\mathbf{n}+\hat{y}) \tau_t(\mathbf{n}+\hat{y}+\hat{y}) 
\tau_y(\mathbf{n}+\hat{y}+\hat{t}) \tau_t(\mathbf{n}+\hat{y}) \nonumber \\
\end{align}
for the $Y$ error (bit flip together with phase flip), where
$\mathbf{n}$ denotes the location of lattice site in 3-dimension
$\mathbf{n} = (i,j,k)$, and $x, y$ denotes the spatial direction and
$t$ denotes the time direction. $\sigma_{x,y,t} (\mathbf{n})$ and
$\tau_{x,y,t} (\mathbf{n})$ represent Ising spin degree of freedoms
associated with the lattice site $(i,j,k)$ for the direction of
$x,y,t$: e.g., $\sigma_x (\mathbf{n})$ lies between the site $(i,j,k)$
and the site $(i+1,j,k)$, and $J$'s denote the couplings. This Hamiltonian has $Z(2) \times Z(2)$ gauge symmetry because it is invariant under $\Lambda_1 \sigma \Lambda_2^{-1}$ and $\Lambda_3 \tau {\Lambda_4}^{-1}$ where local $\Lambda_{1,2,3,4} \in Z(2)$. For example, to study the noise model of independent $XZ$
plus syndrome noise, the two types of variables decouple such that we can remove the $\tau$ spin degrees of freedom
from the Hamiltonian and choose $J_x (\mathbf{n};X) \neq J_y
(\mathbf{n};X)$ as in Eq.~\ref{prob_RPM}. Then, the model becomes a
3-dimensional $Z(2)$ gauge theory with non-uniform couplings for the
plaquettes, which may be called as ``Random Plaquette Gauge Model (RPGM)''. For the symmetric depolarizing noise case of random coupled
plaquette model, we choose $J_x(\mathbf{n};X) = J_y(\mathbf{n};X) =
J_x(\mathbf{n};Y) = J_y(\mathbf{n};Y) = J_x(\mathbf{n};Z) =
J_y(\mathbf{n};Z) = \frac{1}{3} J_t(\mathbf{n};q)$. For the (asymmetric)
depolarizing circuit-noise plus syndrome noise case of random coupled
plaquette model, $J_x(\mathbf{n};X) \neq J_y(\mathbf{n};X),
J_x(\mathbf{n};Z) \neq J_y(\mathbf{n};Z)$ and $J_x(\mathbf{n};Y) = J_y
(\mathbf{n};Y)$ are chosen as in Eq.~\ref{prob_aniso}. In these cases,
the model becomes a 3-dimensional $Z(2) \times Z(2)$ gauge theory with
anisotropic couplings among plaquettes, which may be called the 
``Random Coupled-Plaquette Gauge Model (RCPGM)''.

A Pauli error on a qubit, which is detected by syndrome measurements,
results in the wrong-sign coupling $J_{x,y}$ of the coupling
associated with this qubit and a syndrome error results in the
wrong-sign coupling $J_t$'s. Then, these statistical mechanics models
become variants of quenched disordered spin systems which have many
local minima due to the frustration of couplings. This makes standard
Monte Carlo simulation difficult to converge. To overcome this
problem, we perform parallel tempering steps between Metropolis
updates at a given temperature which involves swapping configurations
from neighboring temperatures using the Boltzmann
weight~\cite{Earl2005}. Schematically, the Monte Carlo simulation proceeds
as: (1) Wrong-sign bond or wrong-sign plaquette interaction
configurations for a given Hamiltonian are drawn with a given quenching
probability. (2) For the given configuration of interaction couplings,
a fixed number of Metropolis steps for the thermalization at each
temperature is taken. (3) For the measurement of observables, a
certain number of runs for Metropolis updates is taken. In-between
these updates, a parallel tempering step \cite{Earl2005} between
neighboring temperatures starting from high temperature is
done. Measurements of the observables are binned in a regular interval
between these combined Monte Carlo updates. The whole process is
repeated over different quenched interaction
configurations. Therefore, two different averages are involved (one
over the thermal ensemble and the other over random configurations of
wrong-signs) and there may be two different phase transitions: one for
thermal transition and the other spin glass transition.

As these spin models are $Z(2)$ gauge theories, the order
parameter for the phase diagram study should be a gauge invariant
observable due to Elitzur's theorem~\cite{Elitzur1975}. Previous studies
based on simpler quantum error models (~\cite{Wang2003},
~\cite{Andrist2010}) use the Wilson loop, $\left< W_C \right> = \left<
\prod_{i\in C} \sigma_i \right>$, where $C$ denotes any closed curve
on the lattice and two different behaviors of the Wilson loop as a
function of the loop size is used to distinguish the phases. Wang et
al.~\cite{Wang2003} considered whether the Wilson loop follows an
area law or a perimeter law to distinguish the phase and studied the
transition at $T = 0$ in detail using the homology of error
chains. Ohno et al.~\cite{Ohno2004} as well as Kubica et
al.~\cite{Kubica2018} investigated the specific heat, in addition to
the Wilson loop behavior. Andrist et al. ~\cite{Andrist2010} studied
the cumulant of the elementary (i.e. smallest area) Wilson loop to
locate the thermal transition temperature. In contrast, we use
the Polyakov line, $P(i,j)=\prod_{t} \sigma_t (i,j)$, as
the order parameter, which is routinely used in studies of Yang-Mills
theory (e.g., ~\cite{Borsanyi2022}) and is closely related to the
Wilson loop. Since a first order thermal transition is
expected~\cite{Andrist2010}, we consider the third order cumulant
together with the susceptibility of the Polyakov line,
\begin{equation}
\langle |\overline{P}| \rangle, \;\; \overline{P} =
\frac{1}{L^2}\sum_{i,j} P (i,j) = \frac{1}{L^2}
\sum_{i,j} \prod_{t} \sigma_t (i,j), \;\;\;\;\; \chi = \langle
\tilde{P}^2 \rangle , \;\;\;\;\; B_3 = \langle \tilde{P}^3 \rangle /
\langle \tilde{P}^2 \rangle^{3/2},
\label{eq:polyakov} 
\end{equation}
where $(i,j)$ denotes the space-like sites and $\prod_t$ means
taking a product along the time-direction at a given $(i,j)$ and
$\tilde{P} = |\overline{P}| - \langle |\overline{P}|\rangle$. Due to periodic boundary conditions, the Polyakov line is
gauge-invariant. Since the Polyakov line in our model is a product of
Ising spin variables, the Polyakov line at $\mathbf x$ itself has
$\pm$-sign and $\langle \overline{P} \rangle$ serves as the "average
magnetization" over the lattice volume and is less susceptible to
short distance fluctuation since the product in Eq.~\ref{eq:polyakov}
is over the entire time direction.

%\textcolor{red}{figures need to updated}
Fig.~\ref{fig:Poly} shows typical behaviors of the average Polyakov
line across different noise probabilities. Without wrong-sign
plaquettes (i.e., $p = 2.88 \times 10^{-5}$), the behavior of the average
Polyakov line shows a well-defined transition temperature in the infinite volume
limit. Well-above the threshold probability (see figures with $p =
0.00852$ in Fig.~\ref{fig:Poly},~\ref{fig:Polysusc},
and~\ref{fig:B3}), the average Polyakov line does not show a
transition as the lattice volume increases. Below and near the
threshold probability ($p = 0.00682$), the order parameter still shows
a transition. The third order cumulant and the susceptibility
corroborate this observation. In Fig.~\ref{fig:B3} and
Fig.~\ref{fig:Polysusc}, $B_3$ crosses zero at the temperature
where $\chi$ reaches a peak. Well above the threshold probability ($p
= 0.00852$), even at low temperature, $B_3$ does not cross zero
and $\chi$ does not reach a peak. Between these two extreme noise
cases, $B_3$ crosses zero and $\chi$ still shows a peak at a similar
temperature. We note that finite volume effects are important near the
threshold probability~\cite{Binder1984}. MC result at $p = 0.020$ (which is not included in Fig.~\ref{fig:Phase}) shows a thermal transition but the transition temperature, $T_c$ from $8^3, 12^3, 16^3, 20^3$ and $24^3$ keep decreasing as the lattice
volume increases and it appears that the infinite volume limit of $T_c$ does not exist.

\section{Result}

In this section, an investigation of the 2-dimensional toric/sufrace code for quantum memory with realistic circuit-level
quantum noise (i.e., bit-flip error and/or phase-flip error on qubits
together with syndrome measurement error) is illustrated as an
example. The statistical mechanics model which corresponds to this
case is 3-dimensional $Z(2) \times Z(2)$ gauge theory with anisotropic
couplings and some of the couplings given in Eq.~\ref{eq:Hamiltonian} have wrong-signs.

\begin{figure}[ht]
\centering
\includegraphics[width=0.6\textwidth]{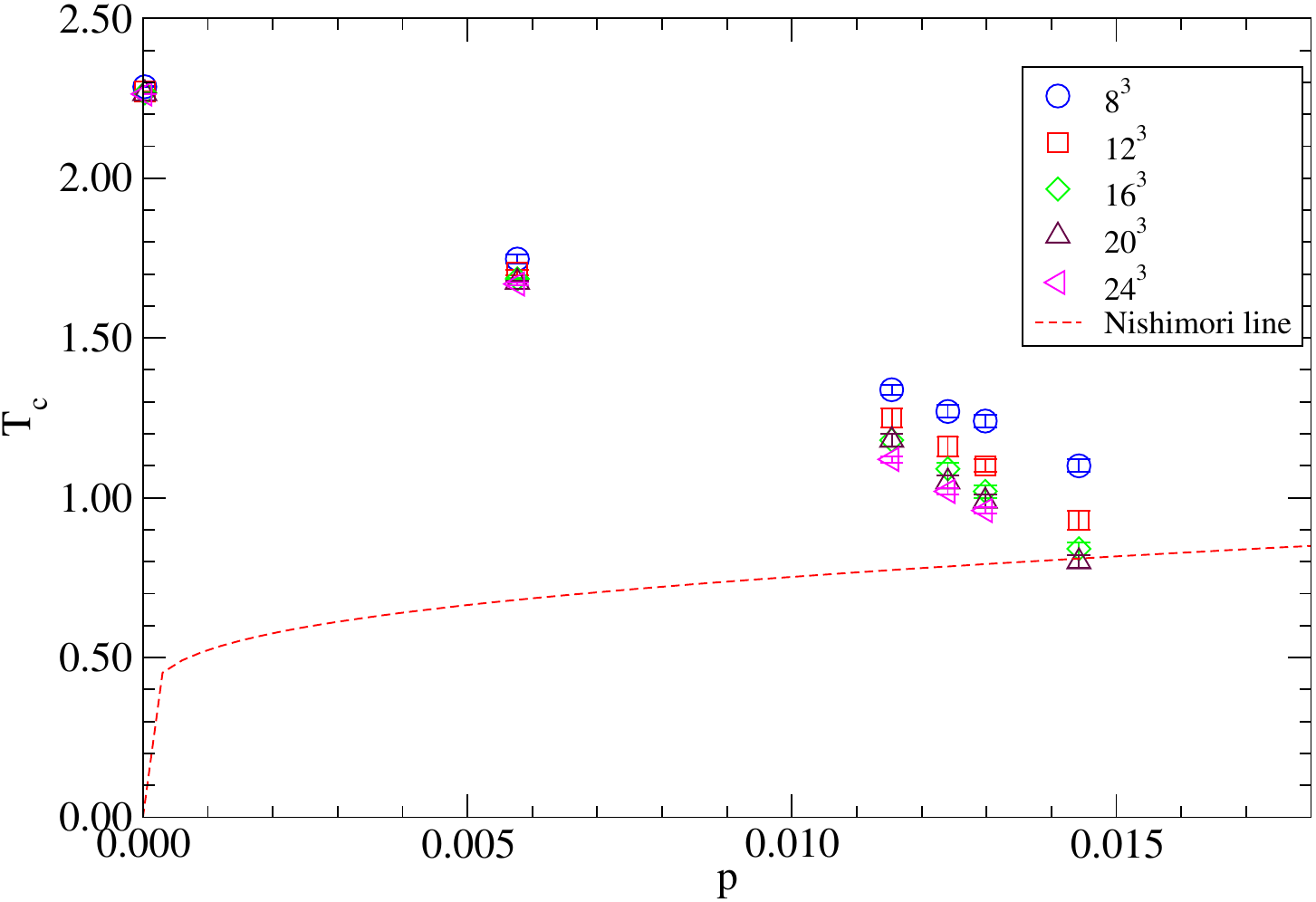}
\caption{Thermal transition temperature vs. wrong-sign probability
on various 3-dimensional lattices for $Z(2) \times Z(2)$ gauge theory
which corresponds to the case of a realistic circuit-noise for bit-flip
error/phase-flip errors together with syndrome measurement errors.} 
\label{fig:Phase}
\end{figure}

Fig.~\ref{fig:Poly} shows the behavior of the Polyakov line for the
case of $p = 2.88 \times 10^{-5}$ (left) which is well below $p_c$
and for $p = 2.31 \times 10^{-2}$ (right) which is well above $p_c$
(Note that we find that the threshold probability, $p_{c} \simeq
0.0144$ in this model from the behavior of various observables). Well
below the threshold probability, the system shows a sudden jump in the
Polyakov line which is in accordance with a first order
transition. With a large quantum error ($p >> p_c$), clearly there is
no transition. The behavior of Polyakov line susceptibility,
Fig.~\ref{fig:Polysusc} and the behavior of the third order cumulant of
Polyakov line, Fig.~\ref{fig:B3} agrees with the observation based on
the Polyakov line results. The transition temperature can be located
by comparing the peak of the susceptibility and the third order
cumulant. The thermal transition temperature is tracked as the the
wrong-sign probability is increased from smaller $p$ to larger
$p$. $T_c$ vs. $p$ is plotted in Fig.~\ref{fig:Phase}.

\section{Conclusion}

Through Monte Carlo simulation (metropolis steps with parallel tempering steps), we investigate phase diagrams of various versions of $Z(2)$ lattice gauge theories (one of which is Random Coupled-Plaquette Gauge Model (RCPGM) having $Z(2) \times Z(2)$ gauge symmetry) in connection with studying the threshold probability for Quantum Error Correction (QEC) protocol of toric/surface code with data qubit error and syndrome measurement error. The susceptibility and the third order cumulant of the Polyakov line allows us to map the phase transitions in the $(T, p)$ space (temperature and the disorder probability) and locate the threshold probability. 

The fault-tolerant implementation of toric/surface code is being hotly pursued (see e.g., the recent work by Google Quantum and AI collaboration~\cite{google_quantum}). The threshold probability is associated with the decoding problem of QEC. The lattice models studied in this work are results of mapping QEC protocols of toric/surface with the cases of uniform depolarizing noise with syndrome errors, circuit-level noise (specifically independent $XZ$ noise plus syndrome noise) and anisotropic asymmetric depolarizing noise plus syndrome noise into corresponding statistical mechanics models.

The case for QEC of toric/surface code with bit-flip noise ($X$ error) plus syndrome measurement error or with phase-flip noise ($Z$ error) plus syndrome noise is mapped into 3-dimensional anisotropic $Z(2)$ gauge theory. Preliminary result of Monte Carlo simulation shows the threshold probability for this case is $p_c \simeq 0.00682$. The uniform depolarizing noise (${\rm pr} (X) = {\rm pr} (Y) = {\rm pr} (Z) = 1/3 p$) (where $Y$-erorr is marginalized) plus syndrome noise (${\rm pr} (q) = p$) case in the toric/surface code is mapped into three-dimensional isotropic $Z(2) \times Z(2)$ gauge theory. Again, preliminary Monte Carlo simulation result suggests the threshold probability of $p_c \simeq 0.06$. The most complex case, $(X, Y, Z)$ data error (where $Y-$ error is marginalized) plus syndrome measurement error is mapped into a three-dimensional anisotropic $Z(2) \times Z(2)$ gauge theory. Preliminary Monte Carlo simulation suggests $p_c \simeq 0.0144$.

Mapping QEC decoding problems under realistic noise models into statistical mechanics models and studying the threshold probability via Monte Carlo simulation is useful tool for realizing fault-tolerant quantum computer. Extending this method to other QEC codes such as color codes will be explored in the future.

\vspace{0.5cm}
{\bf Acknowledgements}

SK is supported by the National Research Foundation of Korea under
grant NRF-2021R1A2C1092701 and in part by NRF-2008-000458 funded by the Korean government and
by the IITP(Institute of Information \& Coummunications Technology
Planning \& Evaluation)-ITRC(Information Technology Research Center)
grant funded by the Korea government(Ministry of Science and
ICT)(IITP-2024-RS-2024-00437191).

\end{document}